\documentclass[12pt]{article}
\usepackage{graphicx}

\newcommand{\Frac}[2]{\frac{\displaystyle #1}{\displaystyle #2}}      
\newcommand{\beq}{\begin{equation}}
\newcommand{\eeq}{\end{equation}}
\newcommand{\beqn}{\begin{eqnarray}}
\newcommand{\eeqn}{\end{eqnarray}}
\newcommand{\beqns}{\begin{eqnarray*}}
\newcommand{\eeqns}{\end{eqnarray*}}

\begin{document}
\begin{titlepage}
\begin{center}
\hfill  INFNNA-IV-2001/24\\
\hfill  DSFNA-IV-2001/24\\
\hfill  CERN-TH/2001-312\\
\hfill  November  2001

\vspace{0.8cm}

{\Large {\bf Long-distance contribution to the muon-polarization 
asymmetry
in $K^+\rightarrow\pi^+ \mu^+\mu^-$ ${}^*$\\}}
\vspace*{1.5cm}
{  Giancarlo D'Ambrosio$^\dagger$ }$\; \; $ and $ \; \; $
{  Dao-Neng Gao$^\ddagger$}
\vspace*{0.4cm} \\ {\it Istituto Nazionale di Fisica Nucleare, Sezione 
di Napoli, Dipartimento di Scienze Fisiche, Universit\`a di Napoli,     
I-80126 Naples, Italy}

\vspace*{1cm} 
\end{center}  
\begin{abstract}
\noindent
We revisit the calculation of the long-distance contribution to the 
muon-polarization asymmetry $\Delta_{LR}$, which arises, in
$K^+\rightarrow\pi^+\mu^+\mu^-$, from the two-photon intermediate 
state.  The parity-violating amplitude of this
process, induced
by the local anomalous $K^+\pi^-\gamma^*\gamma^*$ transition, is
analysed; unfortunately, one cannot
expect to predict its contribution to the asymmetry by using
chiral perturbation theory alone. Here we evaluate this
amplitude and its contribution to $\Delta_{LR}$ by 
employing a phenomenological model called
the FMV model, in which the utility of the vector and axial-vector
resonances exchange is important to soften the ultraviolet behaviour of
the transition. 
We find that the long-distance contribution is of the same order of 
magnitude as the standard model short-distance contribution.
\end{abstract}

\vfill 
\noindent
$^\dagger$ E-mail:~Giancarlo.D'Ambrosio@cern.ch, and on leave of absence 
at {\it Theory Division, CERN, CH-1211 Geneva 23, Switzerland}.\\
$^{\ddagger}$ E-mail:~gao@na.infn.it, and on leave from {\it the
Department of
Astronomy and Applied Physics, University of Science and Technology of
China, Hefei, Anhui 230026, China}.\\
\noindent * Work supported in part by
TMR, EC--Contract No. ERBFMRX-CT980169
(EURODA$\Phi$NE).
\end{titlepage}
\newpage

\section{Introduction}

\par 

The measurement of the muon polarization asymmetry in the decay
$K^+\rightarrow\pi^+\mu^+\mu^-$ is expected to give some  valuable
information on the structure of the weak interactions and flavour
mixing angles \cite{SW90, LWS92, BGT93, BB94, DI98, DV99}.  
The total decay rate for this transition is dominated by the one-photon
exchange contribution, which is parity-conserving, and the
corresponding invariant amplitude  can be parametrized in terms of one
form factor \cite{SW90, DEIP98}: 
\beq\label{PCIA}
{\cal M}^{\rm PC}=\Frac{s_1 G_F\alpha}{\sqrt{2}}f(s) (p_K+p_\pi)^\mu
\bar{u}(p_-,s_-)\gamma_\mu v(p_+,s_+),
\eeq
where $p_K$, $p_\pi$, and $p_\pm$ are the four-momenta of the kaon, pion,
and $\mu^\pm$ respectively,  and $s_1$ is the sine of the
Cabibbo angle. The $s_\pm$ is the spin vectors for the $\mu^\pm$, and the
quantity $s=(p_++p_-)^2$ is the $\mu^+\mu^-$ pair invariant mass squared. 

In the standard model, in addition to the dominant contribution in
eq. (\ref{PCIA}),
the decay amplitude also contains a small
parity-violating piece, which generally has the form \cite{LWS92}
\beq\label{PVIA}
{\cal M}^{\rm PV}=\frac{s_1 G_F\alpha}
{\sqrt{2}}\left[B(p_K+p_\pi)^\mu+C(p_K-p_\pi)^\mu\right]
\bar{u}(p_-,s_-)\gamma_\mu \gamma_5 v(p_+,s_+),
\eeq
where the form factors $B$ and $C$ get contributions from both short- and
long-distance physics. 

The muon-polarization asymmetry in $K^+\rightarrow\pi^+\mu^+\mu^-$ is
defined as
\beq\label{mupol}
\Delta_{LR}=\Frac{|\Gamma_R-\Gamma_L|}{\Gamma_R+\Gamma_L},
\eeq
where $\Gamma_R$ and $\Gamma_L$ are the rates to produce right- and
left-handed $\mu^+$ respectively. This asymmetry arises from the
interference of the parity-conserving part of the decay amplitude
[eq. (\ref{PCIA})] with the
parity-violating part [eq. (\ref{PVIA})], which
gives
\beqn\label{asy1}
\Frac{d(\Gamma_R-\Gamma_L)}{d \cos\theta ds}&=&\Frac{-s_1^2 G_F^2
\alpha^2}{2^8m_K^3
\pi^3}\sqrt{1-\frac{4m_\mu^2}{s}}~\lambda(s,m_K^2,m_\pi^2)\nonumber \\ 
&&\times \left\{{\rm Re}
[f^*(s) B]\sqrt{1-\frac{4m_\mu^2}{s}}\lambda^{1/2}(s,m_K^2,m_\pi^2)
\sin^2\theta\right.\nonumber\\
&&\left. + ~4\left({\rm
Re}[f^*(s) B]\frac{m_K^2-m_\pi^2}{s}
+{\rm Re}[f^*(s) C] \right)m_\mu^2 \cos\theta\right\},
\eeqn
while, as a good approximation, the total decay rate can be obtained from
eq. (\ref{PCIA}): 
\beqn\label{rate}
\Frac{d(\Gamma_R+\Gamma_L)}{d\cos\theta ds}=\Frac{s_1^2
G_F^2 \alpha^2|f(s)|^2} {2^9m_K^3
\pi^3}\sqrt{1-\frac{4m_\mu^2}{s}}~\lambda^{3/2}(s,m_K^2,m_\pi^2)\nonumber\\
\times\left[1-\left(1-\frac{4m_\mu^2}{s}\right)
\cos^2\theta\right],
\eeqn
where $\lambda(a,b,c)=a^2+b^2+c^2-2(ab+ac+bc)$, $4m_\mu^2\le s \le
(m_K-m_\pi)^2$, and $\theta$ is the angle
between the three-momentum of the kaon and the
three-momentum of the $\mu^-$ in the $\mu^+\mu^-$ pair rest frame.
It is easy to see that, when the decay distribution in eq. (\ref{asy1}) is
integrated over $\theta$
on the full phase space, the contribution to $\Delta_{LR}$ from the $C$
part amplitude vanishes. 
 
Fortunately, the form factor $f(s)$ in eq. (\ref{PCIA}) is now known. In 
fact chiral
perturbation theory dictates the following decomposition \cite{DEIP98}: 
\beq\label{form1}
f(s)=a_++b_+ \frac{s}{m_K^2}+w_+^{\pi\pi}(s/m_K^2).
\eeq
Here, $w_+^{\pi\pi}$ denotes the pion-loop contribution, which leads to a 
small 
imaginary part of $f(s)$ \cite{DEIP98, EPR87}, and its full expression   
can be found in Ref. \cite{DEIP98}.  This structure has been accurately
tested and found correct by the E856 Collaboration, which fixes also: 
$a_+=-0.300\pm0.005$, $b_+=-0.335\pm0.022$ \cite{Appel99}. 
Recently the HyperCP Collaboration also studied this channel in 
connection with the study of CP-violating width charge asymmetry in 
$K^\pm\rightarrow\pi^\pm \mu^+ \mu^-$ \cite{HyperCP}. This channel will be 
further analysed by E949 \cite{E949} and NA48b \cite{NA48b}.

It is known that, within the standard model, the short-distance
contributions to ${\cal M}^{\rm PV}$ in eq. (\ref{PVIA}) arise
predominantly from the
$W$-box and $Z$-penguin Feynman diagrams, which carry clean information on
the weak mixing angles \cite{LWS92}. The authors of Ref. \cite{BB94} 
generalized the results in Ref. \cite{LWS92} beyond the leading 
logarithmic approximation: for the Wolfenstein parameter $\rho$ in the 
range 
$-0.25\le\rho \le0.25$, $|V_{cb}|=0.040\pm0.004$, and $m_t=(170\pm 20)$ 
GeV, 
\beq\label{BBmu}
3.0\times 10^{-3}\le\Delta_{LR}\le 9.6\times 10^{-3}
\eeq
with the cut $-0.5\le\cos\theta \le 1.0$.
Hence, the experimental determination of  $B$
and $C$ would be very interesting from the theoretical point of view,
provided that the long-distance contributions are under control.  

\begin{figure}[t]
\begin{center}
\includegraphics[width=12cm,height=4.4cm]{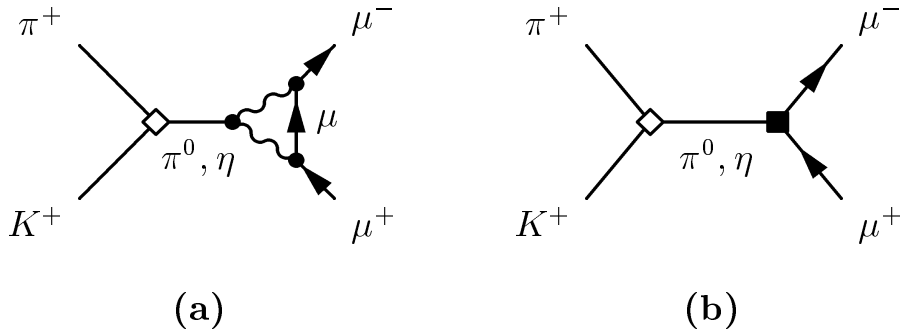} 
\end{center}
\caption{Feynman diagrams that give the two-photon contribution to the
long-distance $C$ part parity-violating amplitude of
$K^+\rightarrow\pi^+\mu^+\mu^-$ in chiral perturbation theory. 
The wavy line is the photon. 
The diamond denotes the weak vertex, the full dot denotes the
strong/electromagnetic vertex, and the full square in (b) denotes 
the local $\pi^0 \mu^+\mu^-$ or $\eta\mu^+\mu^-$ couplings.}
\end{figure}

The dominant long-distance contributions to the parity-violating
amplitude of $K^+\rightarrow\pi^+\mu^+\mu^-$ are from the Feynman diagrams
in which the $\mu^+\mu^-$ pair is produced by two-photon exchange \cite{LWS92}.
Since these contributions arise from non-perturbative QCD, they are
difficult to calculate in a reliable manner. The contribution to the
asymmetry $\Delta_{LR}$ from the long-distance $C$ part 
amplitude, whose Feynman diagrams are shown in Fig. 1,  has been
estimated in Ref. \cite{LWS92} with the cut
$-0.5\le\cos\theta\le 1$, which indicates that it
is substantially smaller than the short-distance part contributions in 
eq. (\ref{BBmu}).  
However, because of the unknown chiral perturbation theory parameters,
the long-distance contribution to $\Delta_{LR}$ from the $B$ part
amplitude, which is induced from the direct $K^+\pi^-\gamma^*\gamma^*$
anomalous transition, has never been predicted. 
As mentioned above, when we integrate over
$\theta$ without any cuts in eq. (\ref{asy1}), only the
contribution from the $B$ part amplitude will survive. Therefore, it would
be interesting to calculate this part of the parity-violating
amplitudes, and estimate its contribution to the asymmetry $\Delta_{LR}$
using phenomenological models.

The paper is organized as follows. In Section 2, we briefly revisit the
two-photon long-distance contributions to $K^+\rightarrow\pi^+\mu^+\mu^-$
within chiral perturbation theory. In order to evaluate the asymmetry
$\Delta_{LR}$ from the long-distance $B$ part amplitude, models are
required. So in Section 3, we will introduce a phenomenological model
involving  vector and axial-vector resonances, called the 
FMV model from Refs. \cite{DP97, DP98}, for this task. 
Section 4 is our conclusions. 

\section{Chiral Perturbation Theory}

In this section we re-examine the two-photon contributions to the
parity-violating amplitude of $K^+\rightarrow\pi^+\mu^+\mu^-$ in 
chiral perturbation theory. There are local terms that can contribute
to the amplitude, which can be constructed using standard notation 
\cite{DI98}.  
The pion and kaon fields are identified as the Goldstone bosons of the
spontaneously broken SU(3)$_L\times$SU(3)$_R$ chiral symmetry and are
collected into a unitary $3\times 3$ matrix $U=u^2={\rm
exp}(i\sqrt{2}\Phi/f_\pi)$ with $f_\pi\simeq$ 93 MeV, and
\beqn\label{Phi}
\Phi=\frac{1}{\sqrt{2}}\lambda\cdot\phi(x)=\left(\begin{array}{ccc}
\frac{\pi^0}{\sqrt{2}}+\frac{\eta_8}{\sqrt{6}} &\; \pi^+ &\;K^+ \\ \; \\
\pi^- &\;-\frac{\pi^0}{\sqrt{2}}+\frac{\eta_8}{\sqrt{6}} &\; K^0\\ \; \\
K^- &\;\bar{K}^0 & \; -\frac{2\eta_8}{\sqrt{6}}\\ \end{array}\right) \ . 
\eeqn
Thus, at the leading order, the local terms contributing to the
decays $K\rightarrow\pi \ell^+\ell^-$ and $K_L\rightarrow\ell^+\ell^-$ can
be written as \cite{LWS92, Val98} 
\beqn\label{local1}
{\cal L}_{\chi}&=&\Frac{is_1
G_F\alpha}{\sqrt{2}}f_\pi^2\bar{\ell}\gamma_\mu\gamma_5\ell\left\{h_1
\langle\lambda_6 Q^2(U\partial^\mu U^+-\partial^\mu U
U^+)\rangle\right.\nonumber \\
&+&h_2 \langle\lambda_6 Q(UQ\partial^\mu U^+-\partial^\mu
UQU^+)\rangle  \nonumber\\
&+& \left. h_3 \langle\lambda_6(UQ^2\partial^\mu
U^+-\partial^\mu U Q^2 U^+)\rangle\right\}, 
\eeqn
where $\langle A\rangle$ denotes Tr($A$) in the flavour space, and $Q$ is
the electromagnetic charge matrix:
\beqn\label{charge}
Q=\left(\begin{array}{ccc}
\frac{2}{3} &\; 0 &\; 0\\ \;\\
0 &\; -\frac{1}{3} &\; 0 \\ \; \\
0 &\; 0&\; -\frac{1}{3}\\ \end{array}\right) \ .
\eeqn

Note that each term contains two $Q$'s because the effective lagrangian
in eq. (\ref{local1}) is from the Feynman diagrams with two photons, and  
$CPS$ symmetry \cite{BDSPW85} has been used to obtain this lagrangian
\cite{LWS92, Val98}. From eqs. (\ref{local1}) and (\ref{PVIA}), it is easy
to obtain the two-photon contributions to the parity-violating form
factors $B$ and $C$ as
\beq\label{chiBC}
B=-\frac{2}{9}(h_1-2h_2+4h_3), \;\;\;\; C=0.
\eeq 

As pointed out in Ref. \cite{LWS92}, $CPS$ symmetry forces the
contribution to $C$ from the leading order local terms to vanish. The
dominant contribution to $C$ is from the $\pi^0 (\eta)$ pole-type
diagrams generated  by the transitions 
$K^+\rightarrow\pi^+\pi^0 (\eta)$ and $\pi^0(\eta)\rightarrow\mu^+\mu^-$,
via the two-photon intermediate states (see Fig. 1). This
contribution to
$\Delta_{LR}$ has been estimated by Lu, Wise, and
Savage \cite{LWS92}: 
$|\Delta_{LR}|<1.2\times 10^{-3}$ for the cut $-0.5\le \cos\theta \le 
1.0$, which is much
less than the asymmetry arising from the short-distance physics in eq. 
(\ref{BBmu}). 

On the other hand, as shown in eq. (\ref{chiBC}), the contribution from
the local terms to the parity-violating form factor $B$ is proportional to 
the constant $h_1-2h_2+4h_3$. 
Since $h_i$'s, $i$ = 1, 2, 3, are unknown coupling constants,
$B$ cannot be predicted in the 
framework of chiral perturbation theory.  
 The lagrangian in eq. (\ref{local1}) also
gives rise to the two-photon contribution to the decay
$K_L\rightarrow\mu^+\mu^-$, which has been studied in Ref. \cite{Val98}
within this context. However, it is not possible to use that decay to
measure the unknown combination $h_1-2h_2+4h_3$ because $h_i$'s in the
local terms enter the amplitude for the transition $K_L\rightarrow
\mu^+\mu^-$ as a different linear combination $h_1+h_2+h_3$.  
Therefore, in the following section, we have to turn our attention to the
phenomenological models and try to estimate this part of the contribution
to the asymmetry $\Delta_{LR}$.

\section{FMV Model}

In order to evaluate the two-photon contribution to the $B$ part amplitude,
one has to construct the local $K^+ \pi^-\gamma\gamma$ coupling that
contains the total antisymmetric tensor. The lowest order contribution
for it starts from $O(p^6)$ in chiral perturbation theory
\cite{EPR88}. Therefore, the unknown couplings in the effective lagrangian
will make it impossible to predict this amplitude, which has been shown in
the previous section. 

Vector meson dominance (VMD) has proved to be very effective in
predicting the coupling constants in the $O(p^4)$ strong lagrangian
\cite{EGPR89, DRV89}; the unknown couplings can thus be reduced
significantly.  However, this is not an easy task in the weak lagrangian
since the weak couplings of spin-1 resonance--pseudoscalar are not yet
fixed by the experiment.  Thus various models implementing weak
interactions at the hadronic level have been proposed, yet it is very
likely that mechanisms and couplings working for a subset of processes 
might not work for other processes unless a secure matching procedure is
provided. Nevertheless the information
provided by the models can be useful to give a general picture of the
hadronization process of the involved dynamics. 

The factorization model (FM) has been widely used in the
literature \cite{EKW93, ENP94, PR91, BEP92} for this task. 
An implementation of the FM in the vector couplings
(FMV) is proposed in Ref. \cite{DP97}; this seems to be  an  
efficient way of
including the $O(p^6)$ vector resonance contributions to
the $K\rightarrow\pi\gamma\gamma$ and
$K_L\rightarrow \gamma\ell^+\ell^-$ processes. The basic statement of FMV 
is to use
the idea of factorization to construct the weak vertices involving the
vector resonances, then integrate out the vectors, i.e. perform the 
factorization at the scale of the vector mass.
An alternative approach is 
to integrate out the vector degrees of freedom to generate the strong 
lagrangian before factorization, i.e. perform the factorization at the 
scale of the kaon mass. As will be shown below, 
the vector and axial-vector resonance degrees of freedom play a very
important role in our
calculation of the one-loop Feynman diagrams of Fig. 2.

Keeping only the relevant terms and assuming nonet symmetry, the strong 
$O(p^3)$ lagrangian linear in the
vector and axial-vector fields 
reads \cite{EGHPR89, Prades94, DP98A}
\beqn\label{strongv}
&&{\cal
L}_V=-\frac{f_V}{2\sqrt{2}}\langle
V_{\mu\nu} f_+^{\mu\nu}\rangle+h_V\epsilon_{\mu\nu\alpha\beta}\langle
V^\mu \{u^\nu, f_+^{\alpha\beta}\}\rangle,\\
\label{stronga}
&&{\cal L}_A=
-\frac{f_A}{2\sqrt{2}}\langle A_{\mu\nu}
f_{-}^{\mu\nu}\rangle+i\alpha_A\langle A_\mu [u_\nu,f_+^{\mu\nu}]\rangle, 
\eeqn 
where
\beqn
u_\mu=iu^+D_\mu U u^+,\\
D_\mu U=\partial_\mu U-i r_\mu U+i U l_\mu,\\
f_\pm^{\mu\nu}=u F_L^{\mu\nu}u^+\pm u^+F_R^{\mu\nu}u,
\eeqn 
$F_{R, L}^{\mu\nu}$ being the strength field tensors associated to the
external $r_\mu$ and $l_\mu$ fields.  If only the electromagnetic field is
considered then $r_\mu=l_\mu=-e Q A_\mu$. We note by
$R_{\mu\nu}=\nabla_\mu R_\nu-\nabla_\nu R_\mu$ ($R=V, A$), and 
$\nabla_\mu$ is the covariant derivative defined as
\beqn
\nabla_\mu R=\partial_\mu R+[\Gamma_\mu, R],\\
\Gamma_\mu=\frac{1}{2}[u^+(\partial_\mu-i r_\mu) u+ u(\partial_\mu-i
l_\mu) u^+].
\eeqn
The determination of the above couplings in eqs. (\ref{strongv}) and
(\ref{stronga}) from the measurements or the
theoretical models has been discussed in Refs. \cite{DP98A, Prades94}.
In order to generate the anomalous $K^+\pi^-\gamma^*\gamma^*$ vertex from 
vector and axial-vector exchange, given the strong  
largangian in eqs. (\ref{strongv}) and (\ref{stronga}), we
have to construct the non-anomalous weak VP$\gamma$ and the anomalous weak
AP$\gamma$ at $O(p^3)$. By applying the factorization procedure with 
the FMV model, we obtain (for details, see the Appendix):
\beqn
{\cal L}_W({\rm VP}\gamma)=-G_8 f_\pi^2 \frac{f_V}{\sqrt{2}}~\eta\langle
\Delta \{V_{\mu\nu}, f_-^{\mu\nu}\}\rangle,\label{VPG}\\
{\cal L}_W({\rm AP}\gamma)=-G_8 f^2_\pi~\ell_A ~\eta
~\epsilon_{\mu\nu\alpha\beta}\langle\{\Delta,
A^\mu\}\{u^\nu,f_+^{\alpha\beta}\}\rangle, \label{APG} 
\eeqn
where $\Delta=u\lambda_6 u^+$, and 
\beq\label{LA}
\ell_A=\frac{3}{16\sqrt{2}\pi^2}f_A\frac{m_A^2}{f_\pi^2}.
\eeq
The $\eta$ is the factorization parameter satisfying $0<\eta\le 1.0$
generally, and it cannot be given by the model.

\begin{figure}[t] 
\begin{center}
\includegraphics[width=13cm,height=5cm]{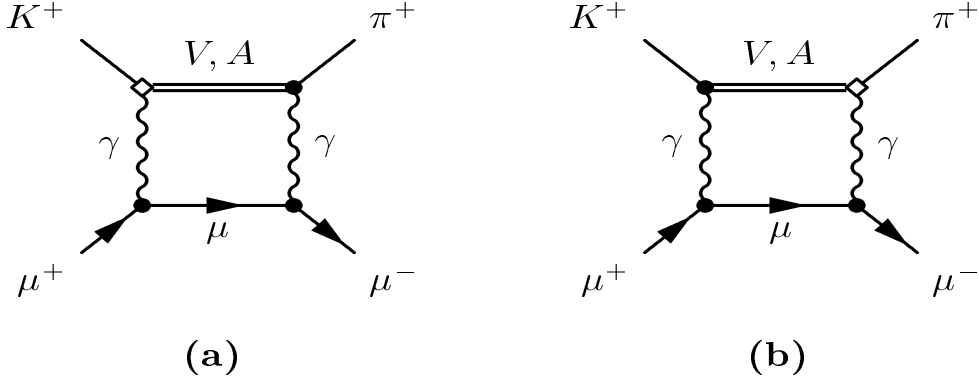}  
\end{center} 
\caption{One-loop Feynman diagrams that give the two-photon
contribution to the long-distance $B$ part parity-violating amplitude of
$K^+\rightarrow\pi^+\mu^+\mu^-$ induced by  the vector and axial-vector
resonances exchange.  The diamond denotes the weak vertex, and the full 
dot denotes the strong/electromagnetic vertex.} 
\end{figure}

Now, combining the strong ${\cal L}_S({\rm VP}\gamma)$ in
eq. (\ref{strongv}) with the weak ${\cal L}_W({\rm VP}\gamma)$ in eq. 
(\ref{VPG}) [or the strong ${\cal
L}_S({\rm AP}\gamma)$ in eq. (\ref{stronga}) with the weak 
${\cal L}_W({\rm AP}\gamma)$ in eq. (\ref{APG})], and 
attaching
the photons to the muons with the usual QED vertices, we can get the
spin-1 resonances contribution to the $B$ part parity-violating amplitude
of $K^+\rightarrow\pi^+\mu^+\mu^-$ from the two-photon intermediate state.
The corresponding Feynman diagrams have been drawn in Fig. 2. 
The calculation of the contribution from the vector resonances
exchange  is straightforward: 
\beqn\label{VIA}
{\cal M}^{\rm PV}_V&=&-i32\sqrt{2}e^4 G_8 f_V h_V
\eta\frac{1}{d}\int\frac{d^4 q}{(2\pi)^4}\frac{1}{(q^2-m_\mu^2)(q^2-m_V^2)}
\nonumber \\
&&\times (p_K+p_\pi)^\mu\bar{u}(p_-,s_-) \gamma_\mu\gamma_5 v(p_+,s_+);
\eeqn
$d$ is the space-time dimension generated from the integral 
$\int d^d q ~q_\mu q_\nu=1/d \int d^d q ~q^2 g_{\mu\nu}$. Because of the
logarithmic divergence in the above equation, we do not set $d=4$ now.
Note that we only retain the loop momentum $q$ in the Feynman
integral of the above equation because  we are concerned about the leading 
order
two-photon contribution to the $B$ part amplitude, and the Feynman
integrals related  to the external momenta are obviously of higher order
with respect to eq. (\ref{VIA}). On the other hand, we do not
consider the diagrams generated by the
weak anomalous $K^+\pi^- V\gamma$ and $K^+\pi^- V V$ couplings, because 
either they have no contributions to the $B$ part amplitude or they are
higher order. 
Likewise, the axial-vector resonance contribution is
\beqn\label{AIA}
{\cal M}^{\rm PV}_A&=&i16\sqrt{2} e^4 G_8 f_A \ell_A
\eta\frac{1}{d}\int\frac{d^4q}
{(2\pi)^4}\frac{1}{(q^2-m_\mu^2)(q^2-m_A^2)}\nonumber\\
&&\times (p_K+p_\pi)^\mu\bar{u}(p_-,s_-) \gamma_\mu
\gamma_5 v(p_+,s_+), 
\eeqn
which receives contributions only from the first term in eq. 
(\ref{stronga}). In fact the contribution from the second term is higher 
order. 

Using the Weinberg sum rules \cite{Wein67} together with the KSRF sum rule
\cite{KSRF66}, one can obtain \cite{EGHPR89}
\beq\label{rules}
f_A=\frac{1}{2}f_V, \;\;\;\; m_A^2=2 m_V^2.
\eeq
Hence, from eq. (\ref{LA}), we have
\beq\label{LAV}
f_A \ell_A=\frac{1}{2}f_V \ell_V,
\eeq
with
\beq\label{LV}
\ell_V=\frac{3}{16\sqrt{2}\pi^2}f_V\frac{m_V^2}{f_\pi^2}. 
\eeq
Then using the relation
\beq\label{LHV}
\ell_V=4 h_V,
\eeq
which, as pointed out in Ref. \cite{DP97},  is exact in the hidden local symmetry
model \cite{BKY88} and also well supported phenomenologically,  we can get
\beqn\label{V+A}
{\cal M}^{\rm PV}_{V+A}&=&-i32\sqrt{2}e^4 G_8 f_V h_V
\eta\frac{1}{d}\int\frac{d^4
q}{(2\pi)^4}\frac{1}{q^2-m_\mu^2}\left[\frac{1}{q^2-m_V^2}
-\frac{1}{q^2-m_A^2}\right]\nonumber\\
&&\times
(p_K+p_\pi)^\mu\bar{u}(p_-,s_-) \gamma_\mu\gamma_5 v(p_+,s_+).  
\eeqn
Fortunately, one will find that the logarithmic divergences in 
eqs. (\ref{VIA}) and (\ref{AIA}) cancel each other, provided that the
relations (\ref{rules}) and (\ref{LHV}) are satisfied. 
Of course, any violation of these relations will lead to the
divergent results; the renormalized procedure is thus needed, and further
uncertainty will be involved. In this paper, we are only concerned about 
the leading order two-photon contribution to the $B$ part amplitude. 
Therefore, 
it is expected that (\ref{rules}) and (\ref{LHV}) could be regarded
as good approximations for this goal. Neglecting 
$m_\mu^2$ in eq. (\ref{V+A}), whose effect is smaller than 5\%, we have
\beq\label{VA}
{\cal M}^{\rm PV}_{V+A}=8\sqrt{2}\alpha^2 G_8 f_V h_V ~\eta ~{\rm
ln}\frac{m_A^2}{m_V^2}(p_K+p_\pi)^\mu \bar{u}(p_-,s_-) \gamma_\mu\gamma_5
v(p_+,s_+).
\eeq
Comparing eq. (\ref{VA}) with eq. (\ref{PVIA}), and using 
\beq
G_8=\frac{s_1 G_F}{\sqrt{2}} g_8,
\eeq  
one will get the two-photon contribution to the parity-violating form
factor $B$ as
\beq
B^{2\gamma}=8\sqrt{2}~\alpha g_8 f_V h_V ~\eta~{\rm
ln}\frac{m_A^2}{m_V^2}.  
\eeq
From the measured $K_S\rightarrow\pi^+\pi^-$ 
decay rate, we fixed $|g_8|$ = 5.1 \cite{EPR87, LWS92}.
Note that $f_V$ and $h_V$ can be determined from the phenomenology of the
vector meson decays, as shown in Ref. \cite{DP98A}, $|f_V|=0.20$, and
$|h_V|=0.037$. Thus, using $m_A^2=2 m_V^2$, we have 
\beq
|B^{2\gamma}|=2.16\times 10^{-3}~\eta.
\eeq                          
Now from eqs. (\ref{mupol}), (\ref{asy1}), and (\ref{rate}), we can get
the asymmetry $\Delta_{LR}$ contributed by $B^{2\gamma}$
\beq
\label{asy10}
\Delta_{LR}=1.7~|B^{2\gamma}|=3.6\times 10^{-3}~\eta
\eeq
for the $\theta$ is integrated over the full phase space, and
\beq
\label{asy11}
\Delta_{LR}=3.0~|B^{2\gamma}|=6.5\times 10^{-3}~\eta
\eeq
for $-0.5\le \cos\theta\le 1.0$.

From eqs. (\ref{asy10}) and (\ref{asy11}), if the factorization
parameter $\eta\simeq 1$, which is implied by the naive factorization,
we find that the long-distance contributions from $B^{2\gamma}$ 
could be compared with the short-distance
contributions given in Refs. \cite{LWS92, BB94}.  In Ref. \cite{DP97},
$\eta\simeq 0.2\sim0.3$ is preferred by fitting the phenomenology of
$K\rightarrow\pi\gamma\gamma$ and $K_L\rightarrow\gamma \ell^+\ell^-$. In
this case, the contributions from eqs. (\ref{asy10}) and
(\ref{asy11}) could be small, though not fully negligible.
Generally, $0<\eta\le 1.0$; therefore, this uncertainty may make it
difficult to get the valuable information on the structures of the weak
interaction and the flavour-mixing angles by the measurement of 
$\Delta_{LR}$ in $K^+\rightarrow\pi^+\mu^+\mu^-$. However 
effects larger than 1\% would be a signal of new physics. 

\section{Conclusions}

We have studied the long-distance contributions via the two-photon
intermediate state to the parity-violating $B$ part amplitude of
$K^+\rightarrow\pi^+\mu^+\mu^-$.  Within chiral perturbation theory,
one can calculate it up to the unknown parameter combination, which is
shown in eq. (\ref{chiBC}). At present, we have no way of estimating
this unknown combination, and it is therefore impossible to determine this
parity-violating amplitude and to predict its contribution to
$\Delta_{LR}$ by using chiral perturbation theory alone.   

We calculate this amplitude in a phenomenological model called the FMV
model. The muon polarization asymmetry $\Delta_{LR}$ has been estimated up
to the factorization parameter $\eta$, which is not given by the
model, but may be in the future determined phenomenologically.
We have established that the background effect may obscure the standard 
model prediction but large new physics effects can still be tested.
\vspace{0.4cm} 

\section*{Acknowledgements}

We would like to thank the CERN TH Division for its nice hospitality and 
Suzy Vascotto for carefully reading the manuscript.
D.N.G. is supported in part by Fellowships from the INFN of Italy 
and the NSF of China. 

\vspace{0.4cm}
\appendix 
\newcounter{pla} 
\renewcommand{\thesection}{\Alph{pla}}
\renewcommand{\theequation}{\Alph{pla}.\arabic{equation}}
\setcounter{pla}{1}
\setcounter{equation}{0}

\section*{Appendix: Non-anomalous weak VP$\gamma$ and 
anomalous weak AP$\gamma$ vertices in FMV}

The $\Delta S=1$ non-leptonic weak interactions are described by an
effective Hamiltonian
\beq
{\cal H}_{\rm eff}^{\Delta S=1}=-\Frac{G_F}{\sqrt{2}}V_{ud}V_{us}^*\sum_i
C_i {\cal Q}_i+ {\rm h.c.}
\eeq
in terms of Wilson coefficients $C_i$ and four-quark local operators
${\cal Q}_i$. If we neglect the penguin contributions, justified by the
$1/N_C$ expansion \cite{PR91}, we can write the octet-dominant piece in
${\cal H}_{\rm eff}^{\Delta S=1}$ as
\beq
{\cal H}_{\rm eff}^{\Delta S=1}=-\Frac{G_F}{2\sqrt{2}}V_{ud}V_{us}^*
C_- {\cal Q}_-+ {\rm h.c.},
\eeq
with 
\beq
{\cal Q}_-=4 (\bar{s}_L\gamma^\mu u_L)(\bar{u}_L\gamma_\mu d_L)-4
(\bar{s}_L\gamma^\mu d_L)(\bar{u}_L\gamma_\mu u_L).
\eeq
The bosonization of the ${\cal Q}_-$ can be carried out in FMV from
the strong action $S$ of a chiral gauge theory. If we split the strong
action and the left-handed current into two pieces: $S=S_1+S_2$ and
${\cal J}_\mu={\cal J}_\mu^1+{\cal J}_\mu^2$, respectively, the ${\cal
Q}_-$ operator is represented, in the factorization approach, by
\beq
{\cal Q}_-\leftrightarrow 4 \left[\langle \lambda \{ {\cal J}_\mu^1,
{\cal J}^\mu_2\}\rangle-\langle \lambda {\cal J}_\mu^1\rangle\langle {\cal
J}^\mu_2\rangle- \langle \lambda {\cal J}_\mu^2\rangle\langle {\cal
J}^\mu_1\rangle \right], 
\eeq
with $\lambda=(\lambda_6-i\lambda_7)/2$; for generality, the currents
have been supposed to have non-zero trace.

In order to apply this procedure to construct the factorizable
contribution to the $O(p^3)$ non-anomalous weak VP$\gamma$ lagrangian, we
have to identify in the full strong action the pieces  that can
contribute at this chiral order. We define, correspondingly,
\beq
S=S_{{\rm V}\gamma}+S_2^{\chi},
\eeq
where $S_{{\rm V}\gamma}$ corresponds to the first term in
eq. (\ref{strongv}), and $S_2^{\chi}$ corresponds to the leading
order [$O(p^2)$] effective lagrangian ${\cal L}_2$ in chiral
perturbation theory for the strong sector.

Evaluating the left-handed currents and keeping only terms of interest we
get
\beqn
\Frac{\delta S_{{\rm V}\gamma}}{\delta
\ell^\mu}&=&-\Frac{f_V}{\sqrt{2}}\nabla^\nu(u^+ V_{\mu\nu}u),\nonumber \\
\Frac{\delta S_2^\chi}{\delta \ell^\mu}&=&-\Frac{f_\pi^2}{2}u^+ u_\mu u.
\eeqn
Then the effective lagrangian in the factorization approach is
\beqn\label{VPG1}
{\cal L}_W^{fact}({\rm VP}\gamma)&=& 4
G_8\eta\left[\left\langle\lambda\left\{\Frac{\delta S_{{\rm
V}\gamma}}{\delta\ell^\mu}, \Frac{\delta S_2^\chi}{\delta
\ell_\mu}\right\}\right\rangle-\left\langle\lambda \Frac{\delta S_{{\rm
V}\gamma}}{\delta\ell^\mu}\right\rangle\left\langle  \Frac{\delta 
S_2^\chi}{\delta
\ell_\mu}\right\rangle \right. \nonumber \\
&& \left. -\left\langle\lambda \Frac{\delta S_2^\chi}{\delta 
\ell^\mu}\right\rangle
\left\langle
\Frac{\delta S_{{\rm 
V}\gamma}}{\delta\ell_\mu}\right\rangle\right]+{\rm h.c.} 
\eeqn  
The explicit term that will give a contribution in our calculations has
been written in eq. (\ref{VPG}).

For the anomalous weak AP$\gamma$ vertex, the corresponding left-handed
currents are
\beqn
\Frac{\delta S_{{\rm A}\gamma}}{\delta
\ell^\mu}&=&-\Frac{f_A}{\sqrt{2}}m_A^2 u^+ A_\mu u,\nonumber \\
\Frac{\delta S_{\rm WZW}}{\delta
\ell_\mu}&=&\epsilon^{\mu\nu\alpha\beta}\Frac{1}{16\pi^2}\left\{F^L_{\nu\alpha}
+\frac{1}{2}U^+ F^R_{\nu\alpha}U, u^+u_\beta u\right\},
\eeqn
where the first term is from the $f_A$ part in eq. (\ref{stronga}), and
the second term is from the Wess--Zumino--Witten lagrangian \cite{WZW}.
An equation similar to eq. (\ref{VPG1}) will be obtained, and the
explicit term for our purpose has been shown in eq. (\ref{APG}).


\begin{thebibliography}{40}
\bibitem{SW90}M. J. Savage and M. B. Wise, Phys. Lett. B {\bf 250}, 151
(1990).
\bibitem{LWS92}M. Lu, M. B. Wise, and M. J. Savage, Phys. Rev. D {\bf 46},
5026 (1992).
\bibitem{BGT93}G. B\'elanger, C. Q. Geng, and P. Turcotte,
Nucl. Phys. {\bf 
B390}, 253 (1993).
\bibitem{BB94}G. Buchalla and A. J. Buras, Phys. Lett. B {\bf 336}, 263
(1994); G. Buchalla, A. J. Buras, and M. E. Lautenbacher, Rev. Mod. Phys. 
{\bf 68}, 1125 (1996). 
\bibitem{DI98}G. D'Ambrosio and G. Isidori, Int. J. Mod. Phys. A
{\bf 13}, 1 (1998). 
\bibitem{DV99}D. G\'omez Dumm and J. Vidal, Phys. Lett. B {\bf 463}, 309
(1999).
\bibitem{DEIP98}G. D'Ambrosio, G. Ecker, G. Isidori, and J. Portol\'es,
JHEP {\bf 08}, 004 (1998).
\bibitem{Appel99}R. Appel {\it et al.}, E865 Collaboration, Phys. Rev.
Lett. {\bf 83}, 4482 (1999).
\bibitem{HyperCP}H. K. Park {\it et al.}, HyperCP Collaboration,
hep-ex/0110033.
\bibitem{E949}L. Littenberg, in the Proceedings of the International 
Conference on CP violation (KAON2001), Pisa, Italy, 12-17 June 2001 and 
hep-ex/0010048.
\bibitem{NA48b}M. Calvetti, in the Proceedings of the 
International Conference on CP violation (KAON2001), Pisa, Italy, 12-17 
June 2001; http://www1.cern.ch/NA48. 
\bibitem{EPR87}G. Ecker, A. Pich, and E. de Rafael, Nucl. Phys. {\bf   
B291}, 692 (1987).
\bibitem{DP97}G. D'Ambrosio and J. Portol\'es, Nucl. Phys. {\bf B492}, 417
(1997).
\bibitem{DP98}G. D'Ambrosio and J. Portol\'es, Nucl. Phys. {\bf B533}, 494 
(1998). 
\bibitem{Val98}G. Valencia, Nucl. Phys. {\bf B517}, 339 (1998).
\bibitem{BDSPW85}C. Bernard, T. Draper, A. Soni, H. D. Politzer, and
M. B. Wise, Phys. Rev. D {\bf 32}, 2343 (1985).  
\bibitem{EPR88}G. Ecker, A. Pich, and E. de Rafael, Nucl. Phys. {\bf
B303}, 665 (1988).
\bibitem{EGPR89}G. Ecker, J. Gasser, A. Pich, and E. de Rafael,
Nucl. Phys. {\bf B321}, 311 (1989).
\bibitem{DRV89}J. F. Donoghue, C. Ramirez, and G. Valencia, Phys. Rev. D
{\bf 39}, 1947 (1989).
\bibitem{EKW93}G. Ecker, J. Kambor, and D. Wyler, Nucl. Phys. {\bf
B394}, 101 (1993).
\bibitem{ENP94}G. Ecker, H. Neufeld, and A. Pich, Nucl. Phys. {\bf B314},  
321 (1994).
\bibitem{PR91}A. Pich and E. de Rafael, Nucl. Phys. {\bf B358}, 311 
(1991). 
\bibitem{BEP92}J. Bijnens, G. Ecker, and A. Pich, Phys. Lett. B {\bf 
286}, 341 (1992). 
\bibitem{EGHPR89}G. Ecker, J. Gasser, H. Leutwyler, A. Pich, and E. de
Rafael, Phys. Lett. B {\bf 223}, 425 (1989).
\bibitem{Prades94}J. Prades, Z. Phys. C {\bf 63}, 491 (1994); {\it 
Erratum},
Eur. Phys. J. C {\bf 11}, 571 (1999). 
\bibitem{DP98A}G. D'Ambrosio and J. Portol\'es, Nucl. Phys. {\bf B533},
523 (1998).
\bibitem{Wein67}S. Weinberg, Phys. Rev. Lett. {\bf 18}, 507 (1967).
\bibitem{KSRF66}K. Karabayashi and M. Suzuki, Phys. Rev. Lett. {\bf 16},
255 (1966); Riazuddin and Fayyazuddin, Phys. Rev. {\bf 147}, 1071 (1966).
\bibitem{BKY88}M. Bando, T. Kugo, and K. Yamawaki, Phys. Rep. {\bf 164}, 
217 (1988).
\bibitem{WZW}J. Wess and B. Zumino, Phys. Lett. {\bf B37}, 95 (1971);
E. Witten, Nucl. Phys. {\bf B223}, 422 (1983).
\end{thebibliography}
\end{document}